\documentclass[fleqn,twoside]{article}
\usepackage{espcrc2}

\usepackage{graphicx}

\newcommand{\be}{\begin{equation}}
\newcommand{\ee}{\end{equation}}
\title{The strangeness content of the nucleon }

\author{UKQCD Collaboration, 
        C.~Michael\address{
 Theoretical Physics Division, Dept. of Math. Sci., \\
          University of Liverpool, Liverpool L69 3BX, UK},
 C. McNeile\addressmark and
 D. Hepburn\address{
 Dept. of Physics, University of Edinburgh, Edinburgh EH9 3JZ, UK}%
 } 

\begin{document}
\begin{abstract}
 We evaluate the matrix element of $\bar{q} q$ in hadron states on a
lattice. We find substantial mixing of the connected and disconnected
contributions so that the lattice result that the disconnected
contribution to the nucleon is large does not imply that the $\bar{s} s$
content is large. This has implications for dark matter searches. 
\vspace{1pc}
\end{abstract} 
\maketitle


\section{INTRODUCTION}

 An important challenge in physics and cosmology is to understand the
nature of dark matter.  One plausible candidate is for this dark matter
to be the lightest supersymmetric  particle: the neutralino. In this
case the dark matter can be detected by scattering from nuclear targets,
 and experimental explorations are currently under way. To extract a
physical  flux from such experiments, one needs the appropriate cross
section for scattering  of a neutralino off a nucleon. This has been
evaluated~\cite{Bottino:1999ei} and depends on  MSSM parameters and on
the QCD matrix elements of the scalar quark current: $ \langle N |
\bar{q} q | N \rangle$.  
 The Higgs exchange terms are dominant and hence the scalar current 
enters multiplied by the relevant quark mass. For this reason, the
strange quark contribution is expected to  be especially important.
Moreover the $u$ and $d$ quark contributions are related to the  $\pi N$
$\sigma$ term and are relatively well known whereas the strange
contribution  is unknown phenomenologically to within a factor of 3. 

 The usual way to parametrise the contribution of the strange quark is by 
 \be
 y = {2 \langle N | \bar{s} s | N \rangle  \over 
        \langle N | \bar{u} u + \bar{d} d | N \rangle }
 \ee
 Estimates~\cite{Borasoy:1996zx} using chiral perturbation theory
suggest $y = 0.2(2) $.

 This ratio of quark matrix elements can be calculated in principle
using lattice methods (see~\cite{Gusken:1999te} for a review). The
required  correlations are illustrated in fig.~1. The disconnected
contribution $D_3$ (numerator of $y$) can either be calculated by 
evaluating the three-point correlator with a disconnected loop or using
the  lattice equivalent of the Feynman-Hellman theorem  to relate it to
the derivative of the  nucleon mass with respect to the sea-quark
hopping parameter (see ref.~\cite{Foster:1998vw} for a discussion  of
this).  Likewise the connected contribution $C_3$ can be evaluated
either as a three-point correlation or as a derivative of the nucleon
mass  with respect to the valence-quark hopping parameter. 
 Then, in this approach we expect $y = D_3/(C_3+D_3)$.

\begin{figure}[htb]
\includegraphics[width=7.5cm]{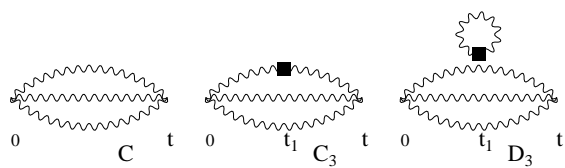}
 \caption{Connected and disconnected diagrams}
 \label{fif.c2}
\end{figure}

 Using $N_f=2$ flavours of sea-quark, SESAM~\cite{Gusken:1998wy} obtain
$y=0.59(13)$ and, using the  same method, we would obtain a similar
result. Using the  three-point correlator approach, it is possible to
estimate $y$ in quenched  studies and results of around 0.6 were
obtained~\cite{Fukugita:1995ba} although the  Kentucky group 
argued~\cite{Dong:1996ec} that renormalisation effects should reduce
this to around  0.36(3) (but see criticism of this approach in
ref~\cite{Gusken:1998wy}).

 This large value of $y$ obtained from lattice studies is surprising 
and it also has major implications for the analysis of dark matter
scattering experiments.  Here we discuss the status of these lattice
determinations critically and we conclude that $y$ is consistent with
zero.

\section{LATTICE ANALYSIS}

 The Feynman-Hellman theorem relates matrix elements of scalar quark
currents  in a nucleon to derivatives of the nucleon mass with respect
to the quark mass. These identities can be derived both in the continuum
and on the lattice.  Here we consider Wilson-like
lattice fermion formulations. 
 The lattice equivalent of the Feynman-Hellman theorem is
that~\cite{Foster:1998vw} the following lattice observables (here we
define $m^b \equiv 1/(2 \kappa)$) are related:

 \begin{equation}
\frac{ \partial (aM_N)} { \partial m_{val}^b }
= 
\lim_{ t_1 , (t-t_1) \rightarrow \infty} 
\frac{ C_3(t_1,t) }{ C(t) }
 \end{equation}

 \begin{equation}
\frac{ \partial (aM_N)} { \partial m_{sea}^b }
= - N_f
\lim_{ t_1 , (t-t_1) \rightarrow \infty} 
\frac{ D_3(t_1,t) }{ C(t) }
 \end{equation}

The  input quark mass parameters  can be written in terms of the
physical bare quark masses, with  an additive mass renormalisation
($m_A$).

\be
m_{val}^{b}  =  m_{A} + a m_{val} \label{eq.val} 
\ee
\be
m_{sea}^{b}  =  m_{A} + a m_{sea} \label{eq.sea}
\ee

 Now  $m_A$ may be determined by varying the valence quark mass
and  determining the critical hopping parameter at which the pion mass
(and hence $a m_{val}$)  becomes zero, or equivalently by finding the
critical valence hopping parameter at which the PCAC mass becomes zero. 
These extrapolations to determine $m_A$ are at fixed
$m_{sea}^b$ and  hence $m_A$ will depend on $m_{sea}^b$. 
 The lattice spacing $a$ also depends on the bare sea-quark 
mass parameter $m_{sea}^b$. 
 These effects can be summarised by the following derivatives
 \be 
 X={d m_A \over d m_{sea}^b}
 {\rm \ \ \ \ and \ \ \ \ \ }
 B = {d \log a \over d m_{sea}^b}
 \ee

We are interested in the  disconnected  scalar matrix element which is
related in the continuum to the derivative with  respect to the sea
quark mass.  On a lattice this   disconnected scalar matrix element is
related to the derivative of the  nucleon mass with respect to the
sea-quark mass  parameter at fixed  valence-quark hopping parameter and
fixed $\beta$ by eq.~\ref{eq.sea}. The key observation is then that as
the sea-quark hopping parameter is varied on the lattice, both  the
valence quark mass and the lattice spacing also change. Thus one needs
to  correct for these  changes to obtain the derivative of the nucleon
mass at fixed valence-quark mass which  is required.

 For  the valence dependence, the required derivative is 
directly given by the lattice observable
 \be
\frac{ \partial M_N} { \partial m_{val} }
=\frac{ \partial (aM_N)} { \partial m_{val}^b }
 \ee
 For the sea-quark derivative, however, there will be several other
factors involved as discussed above:
 \begin{eqnarray}
\frac{ \partial M_N} { \partial m_{sea} } (1- a m_{sea} B - X)  
=   \ \ \ \ \ \  & & \nonumber \\
  \frac{ \partial (aM_N)} { \partial m_{sea}^b }  
 + ( a m_{val} B + X)\frac{ \partial (aM_N)} { \partial m_{val}^b } 
 - M_N B & &
 \label{eq.full}
 \end{eqnarray}

 This shows that the lattice connected and disconnected contributions
are mixed  when related to the more physical derivatives at fixed bare
quark mass and  fixed scale $a$. There will  also be additional
perturbative matching  contributions to take into account to have a
precise link between lattice observables and the continuum expressions.,
but we do not discuss these further here.

 We evaluate the above expressions using UKQCD
data~\cite{Allton:1998gi,Allton:2001sk}. We mainly use data from 
$\beta=5.2$ on $16^3 32$ lattices with $N_f=2$ flavours of sea quark
with $\kappa=0.1355$ or $0.1350$ and using a  NP-clover formalism with
$C_{SW}=2.0171$. These hopping parameter values correspond to quark masses
around the strange quark (since the  $\pi / \rho$ ratio is 0.58 and
0.70 respectively). From the $r_0$ values~\cite{Allton:2001sk}, we obtain
 $B=4.4(8)$ while from extrapolating the PCAC masses to obtain the
critical hopping parameters  at these two sea quark masses we obtain
$X=-0.66(4)$. This implies that there will be substantial mixing of the
lattice disconnected  and connected contributions in evaluating the
derivative with respect to the sea-quark mass. 

 Setting $B=X=0$ in eq.~\ref{eq.full} gives the naive lattice ratio of
$y=0.53(12)$  while including the full mixing gives $y=-0.28(33)$.

 Since we are using finite differences to evaluate the derivatives in 
eq.~\ref{eq.full}, we may instead use the  value of  the nucleon mass in
physical  units for the four cases ($\kappa_{sea}=0.1355,\ 0.1350$; 
$\kappa_{val}=0.1355,\ 0.1350$) and evaluate the quark mass in each case
from  the bare quark mass (eqs.~{\ref{eq.val},\ref{eq.sea}). We prefer
to use the  bare quark mass here because of the lattice Feynman-Hellman
theorem, but it would be possible to use the lattice PCAC quark mass,
extrapolate to the continuum and  then use the Feynman-Hellman relation in
the continuum. The four combinations of sea and valence quark mass will 
then not lie at the corners of a rectangle, but at corners of some
quadrilateral.  We can then use the nucleon masses at these four values
to evaluate the required derivatives with respect  to the sea quark mass
at constant valence mass and vice versa. The result from  this approach
is $y=-0.30(34)$ which is consistent with the value quoted above. 

 Note that  $y$ is expected to be positive, so the negative value
obtained  is just a reflection of the large statistical error. We have
also attempted to  measure the disconnected lattice correlator using a
three point function approach. The  result was compatible with using
derivatives of the nucleon mass  but the statistical errors were larger.
 
\section{DISCUSSION}

 To obtain a reliable lattice determination of $y$, one should use 
$N_f=2$ flavours of light sea quark plus a heavier (strange mass) sea
quark.  Then the required disconnected diagram can be obtained either as
a derivative of the nucleon mass  with respect to the strange sea quark
mass or by evaluating the $D_3$  diagram with strange quarks in the
disconnected loop. In practice one will need to extrapolate the sea and
valence masses  to  to the physical $u$ and $d$ masses.  This 
extrapolation is known to be non-trivial for the  extraction of the $\pi
N$ sigma term~\cite{Leinweber:2000sa}. A continuum limit of the lattice
observables  should also be taken as well as building in perturbative
matching.

 Instead we use $N_f=2$ flavours of sea-quark which should be a good
approximation for the sea. Extracting the disconnected diagram as a
derivative is only possible if the quarks considered in the disconnected
loop are the sea-quarks.  It is possible to go beyond this by explicitly
evaluating $D_3$ with different mass  quarks in the disconnected loop
and this is in progress.  We are also exploring ways to  reduce the 
large statistical errors we find.

 Our main conclusion is that the current lattice data are unable to 
give a determination with any precision of $y$, but that the naive
lattice  ratio of $y \approx 0.6$ is not appropriate and the lattice
result  is indeed compatible with $y=0$.  That $y$ is compatible with
zero  implies that there is no evidence for any dependence of the 
nucleon mass  on the sea quark mass  and this conclusion was also 
reached in an earlier lattice study~\cite{Foster:1998vw}  of the sea
quark dependence of the meson (pseudoscalar and vector) masses.



\end{document}